# Recognition of Words from the EEG Laplacian

**J. Acacio de Barros (a), C. G. Carvalhaes (b), J. P. R. F. de Mendonça (a) and P. Suppes (c)**

(a) Professor Adjunto do Departamento de Física – ICE, Universidade Federal de Juiz de Fora

Juiz de Fora, MG 36033-330, Brazil

(b) Professor Adjunto do Instituto de Matemática e Estatística, Universidade Estadual do Rio de Janeiro

Rio de Janeiro, RJ 20550-900, Brazil

(c) Lucie Stern Professor of Philosophy, Emeritus - CSLI – Stanford University

Stanford, CA 94305-4115, USA

**Abstract**

Recent works on the relationship between the electro-encephalogram (EEG) data and psychological stimuli show that EEG recordings can be used to recognize an auditory stimulus presented to a subject. The recognition rate is, however, strongly affected by technical and physiological artifacts. In this work, subjects were presented seven auditory simuli in the form of English words (*first, second, third, left, right, yes,* and *no*), and the time-locked electric field was recorded with a 64 channel Neuroscan EEG system. We used the surface Laplacian operator to eliminate artifacts due to sources located at regions far from the electrode. Our intent with the Laplacian was to improve the recognition rates of auditory stimuli from the electric field. To compute the Laplacian, we used a spline interpolation from spherical harmonics. The EEG Laplacian of the electric field were average over trials for the same auditory stimulus, and with those averages we constructed prototypes and test samples. In addition to the Laplacian, we applied Butterworth bandpass digital filters to the averaged prototypes and test samples, and compared the filtered test samples against the prototypes using a least squares metric in the time domain. We also analyzed the effects of the spline interpolation order and bandpass filter parameters in the recognition rates. Our results show that the use of the Laplacian improves the recognition rates and suggests a spatial isomorphism between both subjects.

# Introduction

Electroencephalogram (EEG) has been used widely to study processing of words by the brain. Evoked potential components, like the P300 or the N400, indicate collective macroscopic behavior of neurons that is reflected in the EEG as speech is processed. Recently, in a series of papers, Suppes and collaborators (Suppes *et al.*, 1997, 1998, 1999a,b; Suppes and Han, 2000) showed that electric fields recorded via EEG (and, magnetic fields, via MEG – magnetoencephalogram) could be used to recognize auditory stimuli presented to a subject. These results indicate an isomorphism between EEG recorded brain signals and speech, therefore showing that some information about brain processing of speech is available at the field level. The recognition rates of the auditory stimuli from the electric fields obtained by Suppes *et al.* (1997) were several standard deviations above chance recognition, but there is room for improvement, as their rates of correct recognition ranged from 37% to 97%. Improving such recognition rates is important, not only to understand the processing of information inside the brain, but also to be usable in practical applications. For example, the use of such electric fields to control computers and equipment for disabled people would require reliable rates of recognition before they could be used in commercial products.

Possible causes for the misrecognition of words in the electric fields are technical and physiological artifacts. For example, when a subject blinks, disturbances in the electric potential generated at the eye's muscles propagate to the scalp, affecting the EEG signal. In this paper, we use the Laplacian to emphasize the electric ativities that are close to a recording electrode, filtering out those that might have an origin outside of the skull (Nunez, 1981; Nunez and Westdorp, 1994; Infantosi and Almeida, 1990)**.** Our criterion for testing whether the Laplacian removes undesirable signals is the recognition rate of brain representations of spoken words.

# Laplacian Computation

As we mentioned earlier, some sources of noise in the EEG recordings of the scalp potential are bioelectric or other potentials generated outside of the scalp that propagate to the measuring electrode. For example, the generation of static electricity by the subject, the induction of currents from powerlines, and biopotentials that propagate from the cardiac muscle or from other muscles, are all sources that interfere with the measurements of the scalp potential produced by sources within the brain. In order to reduce these noises, it would be interesting to obtain, from the EEG, local information about the brain activity. One possibility is the surface Laplacian of the scalp potential, defined as

$$\nabla^2_{surface} = \nabla_{surface} \cdot \nabla_{surface}, \tag{1}$$

where, in cartesian coordinates, using $\hat{z}$ as the unit basis vector perpendicular to the surface,

$$\nabla_{surface} = \frac{\partial}{\partial x}\hat{x} + \frac{\partial}{\partial y}\hat{y}. \tag{2}$$

The surface Laplacian of the scalp potential is a local operator with a simple physical meaning. Since the Laplacian of the electric potential is the divergent of the electric field, if we assume that the scalp is an ohmic conductor with conductance σ, then the current density **J** is linearly related to the gradient of the potential Φ by

$$\mathbf{J} = -\sigma \nabla \Phi, \tag{3}$$

where

$$\nabla = \frac{\partial}{\partial x}\hat{x} + \frac{\partial}{\partial y}\hat{y} + \frac{\partial}{\partial z}\hat{z}. \tag{4}$$

Furthermore, if we assume that there are no sources of charge at the scalp but there are charges from the skull flowing to the scalp, the three-dimensional divergent of **J** is zero, but the surface Laplacian is not, as, from equation (3) and the definition of the surface Laplacian we have

$$\nabla_{surface} \cdot \mathbf{J} = -\sigma \nabla^2_{surface} \Phi. \tag{5}$$

Thus, the surface Laplacian of the scalp potential is proportional to the local-flux of electric charge from the skull to the scalp. The surface Laplacian, therefore, is a locally measurable property of the brain electrical activity.

One of the practical problems to be solved in using the Laplacian for the recognition of speech using the electric fields is how to compute the second derivative of such a function, if only a finite number of experimental data points is available.[1] The simplest way is to use a discrete approximation in a two-dimensional surface grid. To compute the Laplacian at position $\mathbf{r}_C$ we need to compute the second-order derivative of the potential $\Phi$ on $\mathbf{r}_C$ with respect to $x$ and $y$, the two coordinates defined to be orthogonal and parallel to the surface. We start with the discrete grid shown in Figure 1. Using the method of finite differences (de Moura, 2002; Moin, 2001), it is easy to show that the second derivatives of $\Phi$ with respect to $x$ and $y$ are, at the point $r_0$, approximated by

$$\frac{\partial^2 \Phi}{\partial x^2} \approx \frac{\Phi(\mathbf{r}_1) + \Phi(\mathbf{r}_2) - 2\Phi(\mathbf{r}_0)}{\Delta x^2} + O(\Delta x^2),$$

and similarly for $y$, where $\Phi(\mathbf{r}_i)$ is the potential at point $i$, $i = 0, 1, 2, 3, 4$, and $\Delta x$ ($\Delta y$) is the distance between detectors in the direction $x$ ($y$). Thus, the surface Laplacian of the scalp potential is computed at detector "0" (see Figure 1) as being simply

$$\nabla^2_{surface} \Phi \approx \frac{\Phi(\mathbf{r}_1) + \Phi(\mathbf{r}_2) + \Phi(\mathbf{r}_3) + \Phi(\mathbf{r}_4) - 4\Phi(\mathbf{r}_0)}{\Delta l^2}, \tag{6}$$

where we used the approximation that $\Delta x = \Delta y \equiv \Delta l$. One of the problems with the discrete approximation is the error, due to the finite-size step, particularly, is high if $\Phi$ varies too much within the distance $\Delta l$. This is especially relevant if the source of interest is located between two electrodes.

---

[1] In our case, only 57 out of 64 Neuroscan channels were used to collect data, as 7 channels were used as auxiliary channels.

To address the issue of having a discrete amount of information to compute an essentially continuous function (the Laplacian), it is useful to make continuity and smoothness assumptions on the functions and use interpolation techniques. It is natural, therefore, to ask what kind of interpolation we can use to compute the surface Laplacian of $\Phi$. A good candidate is the spherical spline interpolation, introduced by Perrin *et al.* (1987, 1989), as it yields results that are easily computed numerically.

When doing an interpolation, we want to find out the best function, $\Phi(\mathbf{r})$, that fits the finite set of data points, i.e., the points where we know the actual value of $\Phi$. In other words, given $n$ experimental values of $\Phi(\mathbf{r})$ at electrode positions $\mathbf{r}_i$ ($i = 1, \cdots, n$), we want to find the continuous function $\Phi(\mathbf{r})$ that best fits those values. There are several different ways to interpolate $\Phi(\mathbf{r})$, but for simplicity in our computations we start with the assumption that the potential at point $\mathbf{r}$ can be represented as a supperposition of spherical harmonics as

$$\Phi(\mathbf{r}) = c_0 + \sum_{i=1}^{N} c_i g_m \left( \frac{\mathbf{r}_i \cdot \mathbf{r}}{r_i r} \right), \tag{7}$$

where

$$g_m(x) = \frac{1}{4\pi} \sum_{n=1}^{\infty} \frac{2n+1}{n^m (n+1)^m} P_n(x), \tag{8}$$

$P_n(x)$ is a legendre polynomial on $x$ of order $n$, and $m > 1$ is an integer called the *interpolation order*. We need to find the values of the coefficients $c_i$ that best fit the experimental data. Let us define the following matrices.

$$C = \text{diag}(c_1, ..., c_n),$$

$$Z = \text{diag}(\Phi(\mathbf{r}_1), ..., \Phi(\mathbf{r}_n)),$$

$$G = \begin{pmatrix} g_{11} & g_{12} & \cdots & g_{1n} \\ g_{21} & g_{22} & & g_{2n} \\ \vdots & & \ddots & \vdots \\ g_{n1} & g_{n2} & \cdots & g_{nn} \end{pmatrix},$$

where $g_{ij} = g_m(\mathbf{r}_i \cdot \mathbf{r}/r_i r)$. It follows at once that the coefficients that better fit the data can be written in terms of the above matrices as

$$C = G^{-1}Z. \tag{9}$$

We are setting $c_0 = 0$ for two reasons. First, because $c_0$ is a global value added to a potential, and therefore in most cases irrelevant. Second, since we are interested in computing the derivative of the potential, this term will be thrown away regardless of its value. This reduces the problem to, computationally, finding the inverse to the matrix $G$ and multiplying it by $Z$.

The reason for choosing (7) for the potential is that, once we have the coefficients $c_i$, it is straightforward to obtain the surface Laplacian. This is true because

$$\nabla^2_{surface} P_n = -(2n+1)P_n.$$

Therefore, we obtain at once (Perrin *et al.,* 1989) that

$$\nabla^2_{surface} \Phi(\mathbf{r}) = \sum_{i=1}^{N} c_i h_m\left(\frac{\mathbf{r}_i \cdot \mathbf{r}}{r_i r}\right), \tag{10}$$

where

$$h_m(x) = -\frac{1}{4\pi} \sum_{n=1}^{\infty} \frac{(2n+1)^2}{n^m (n+1)^m} P_n(x). \tag{11}$$

Equation (10), together with (9), is the expression we use to implement the Laplacian of the potential in our computations.

# Experiment and Data Processing

## Data Acquisition

The EEG data-set we used was collected by Suppes and collaborators. The data acquisition was described in details in Suppes *et al.* (1997), and here we will only reproduce the relevant information. For the data we used, corresponding to subjects S6 and S7 in Suppes *et al.* (1997), electric fields were recorded using a 64-channel NeuroScan EEG system at the Palo Alto Veterans Affairs Health Care System. Our choice of subjects S6 and S7 was necessary, as they were the only subjects in the experiment that had their electric fields recorded, through the associated electric potential, with a 64-channel EEG system, since all other subjects used the international 10-20 electrode placement system. Subjects S6 and S7 were normal males, 75 and 30 years-old, respectively. S6 was a native speaker of English, and S7 a native speaker of Chinese, but fluent in English. The electric potentials were recorded with reference to the linked earlobe electrodes. The data was filtered by a bandpass filter ranging from DC to 200 Hz, and sampled at a rate of 500 Hz. Auditory stimuli, with durations of about 300 ms each, were presented to the subjects in intervals that varied from 1.5 to 1.7 s. The stimuli consisted of the seven English words: *first, second, third, yes, no, left,* and *right.* Words were presented in a random order, in a total of approximately 100 trials for each word. Subjects were instructed to listen to the words carefully.

## Data Analysis

To compare the EEG surface potential to the Laplacian, we followed a procedure similar to the one described in Suppes *et al.* (1997), but we used the Laplacian computed at each electrode in *lieu* of the scalp potential. For each such potential a baseline was set by averaging the first 204 observations before the onset of stimulus, and then subtracting this average from each trial. Then we split the data into two sets, one with trials labeled even (E) and another with odd (O), and a prototype was created averaging all trials of a given word in one of the half sets, e.g. the E set. With the remaining trials, test samples were averaged with *n* trials for each word. After prototypes and test samples were created, they were compared to each other via least square distances between the two averaged wave-forms. A test sample was correctly classified from the surface potential if its least square distance to the correct prototype was the smallest one. In order to achieve the best recognition rate, we ran the above procedure for several different bandpass ranges of a butterworth digital filter, in an attempt to filter out artifacts. To filter the data, we computed the FFT of each test sample and prototype using FFTW 3.0 (Frigo and Johnson, 1998). We then applied to the FFTs a fourth-order Butterworh filter and did an inverse FFT, returning to the time domain. The classification scheme used for the scalp potential is the same used by Suppes *et al.* (1997).

To classify using the Laplacian, we first computed the Laplacian for each electrode via spherical spline interpolation of order $m$, and with this new signal, we followed a similar procedure to that described in the previous paragraph. Finally, we made a search for the best classification filter and Laplacian interpolation $m$, searching what combination of values of the filter and $m$ yield the best recognition rate.

### **Results and Discussions**

Our main results are summarized in Table 1. We found improvements in the recognition rates by using the Laplacian. For example, for subject S6 the best recognition rate using the Laplacian was 88% for subject S6, with a bandpass filter from 4.5 Hz to 6.5 Hz, whereas the best recognition rate using the scalp potential was 79%, with a bandpass filter from 2.0 Hz to 12.5 Hz. On the other hand, our worst Laplacian recognition rate was 62% for subject S7 in the composing scheme where the odd trials were used to create the prototypes and the even trials to create the testing samples, and it equalled that of the scalp potential recognition rate. Thus, for this small set of subjects we got consistently better results if we used the Laplacian for our classification scheme. For completeness, we show in Figure 2, the best rates of recognition from the electric fields for different values of the bandpass filter. We can see that the Laplacian narrows the region of good recognition, if compared to the scalp potential.

We also computed the recognition rate for different values of $n$, the number of average trials per test sample. Figure 3 shows the recognition rates as a function of $n$ for the Laplacian and potential. We can see in Figure 3, from linear regression lines, that the Laplacian (solid line) consistently outperforms the potential (dashed line), but both yield approximately the same result with single trials. Even though Figure 3 shows only results for subject S6, both subjects showed the same behavior, with the Laplacian consistently outperforming the potential.

Since the Laplacian is a second-order spatial derivative, and therefore local, it is interesting to look at the spatial distribution of brainwave recognition rates on the scalp. Figure 4 shows the distribution for subjects S6 and S7, both for the Laplacian and the potential. We can see that for the scalp potential, S6's best electrode position is almost opposed to that of subject S7, with both best electrodes being a global maxima. On the other hand, the Laplacian maps show three distinct local maxima for subjects S6 and S7 that have reasonably good recognition rates. Furthermore, we can see that for both subjects the localization of those three regions seem to be similar. These data indicate that the best recognition loci could be invariant among subjects, contrary to the scalp potential.

We also investigated the effects on the recognition rates of using the international 10-20 system of electrode placement instead of the 64-elecrodes. This is an important question, as 10-20 systems are widely available and significantly less expensive. For subject S7, using only electrodes of the 10-20 system we got 79% as our best rate for the scalp potential, using electrode C3 with scheme OE. For the same subject and the same scheme, the Laplacian resulted in a recognition rate of 69%, inferior to the scalp potential. For subject S6, both the Laplacian and the scalp potential resulted in the same rates for the 10-20 system, namely 59%. The spacial distribution of best recognition rates for the Laplacian using a 10-20 system are shown in Figure 5.

## **Conclusions**

We analyzed the experiment described by Suppes *et al.* (1997) using a similar least-squares and filter search procedure with the Laplacian of the potential. The Laplacian was computed using a spherical harmonic spline interpolation method. Our case study indicates that, for the two subjects studied, the Laplacian might give better recognition rates than the original EEG signal. It also indicates that this better result is maintained if we decrease the number of trials used to build the test sample.

In addition to better recognition rates, our Laplacian data suggest a possible important invariance result between subjects that is not present in the scalp potential. If we plot a distribution of recognition rates on the scalp, the Laplacian gives three loci of local maxima, and these three loci seem to be invariant between subjects S6 and S7. These points of local maxima, located at the centroparietal region, also seem to be consistent with recent fMRI (Binder *et al.*, 1997) and older neurophysiological results (Kandel and Schwartz, 1985) for speech processing areas.

Since we only have data available from two subjects, more data should be collected and analyzed to verify if the above results are robust. Unfortunately, the data collected by Suppes *et al.* (1997) used a 10-20 EEG system for most of the subjects, except for subjects S6 and S7. In order to use a smaller number of detectors in the Laplacian computation, we would need to confirm if the existence of the three loci of local minima remains when we downgrade from a 64 channel EEG to a 10-20 one. Figure 5 seems to indicate that they do not, i.e., 10-20 systems are not detailed enough to map the three local maxima. Since the Laplacian is a measurement of local surface current, the invariant loci of maxima suggest the existence of sources of information giving good recognition rates that should be independent. This points to the possibility of using combinations of the Laplacian at these three points to improve the recognition rates obtained.

*Figures and Tables*

# Figures and Tables

**Table 1**: Highest recognition rates for each subject using the potential and Laplacian processing. Shadowed lines correspond to the Laplacian processing. The EO (OE) scheme refers to the analysis in which even (odd) trials are used to compose the prototypes and the odd (even) are used to compose the test samples.

| Subject | Composing Scheme | Highest recognition rate (%) | Parameters for the best results | | | |
|---|---|---|---|---|---|---|
| | | | Best EEG sensor | Best interpolation order ($m$) | Best filter (Hz) | |
| | | | | | low freq. | high freq. |
| S6 | EO | 79 | F5 | 4 | 5.5 | 6.5 |
| | | 76 | C2A | | 5.5 | 6.5 |
| | OE | 88 | C4A | 3 | 4.5 | 6.5 |
| | | 79 | C3 | | 2 | 12.5 |
| S7 | EO | 66 | C1P | 4 | 1 | 4.5 |
| | | 62 | C6A | | 2.5 | 10 |
| | OE | 62 | F7 | 4 | 1.5 | 22.5 |
| | | 62 | C6A | | 2.5 | 8 |

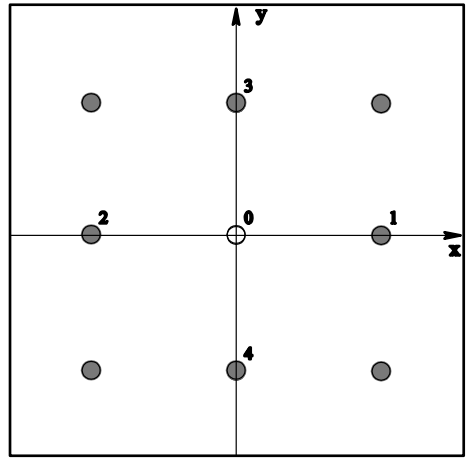

**Figure 1:** Grid showing nine EEG detectors and the coordinate system parallel to the surface, *x* and *y* . In this Figure we choose the origin of the coordinate system to coincide with the central electrode, labeled "0", whose coordinates are (0,0). Each small circle represents the position of an actual EEG electrode.

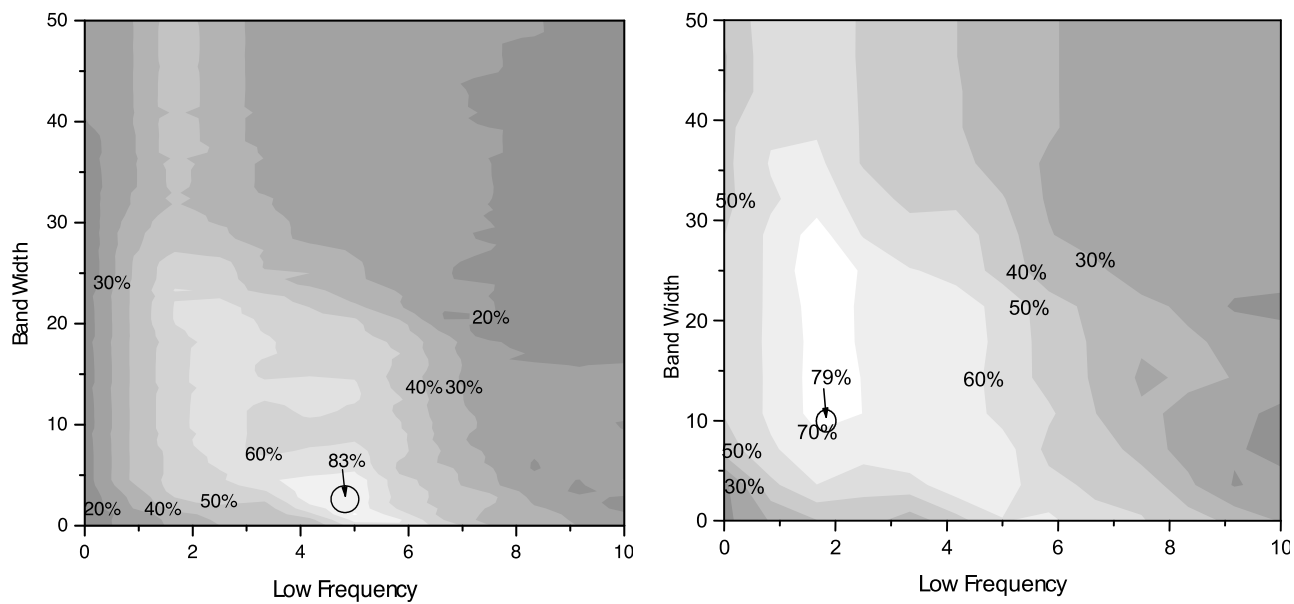

**Figure 2:** Distribution of recognition rates, for subject S6, with different filters. The left graph shows electrode C4A and the recognition was computed using the Laplacian. The right graph shows the recognition rates of using the potential at electrode C3.

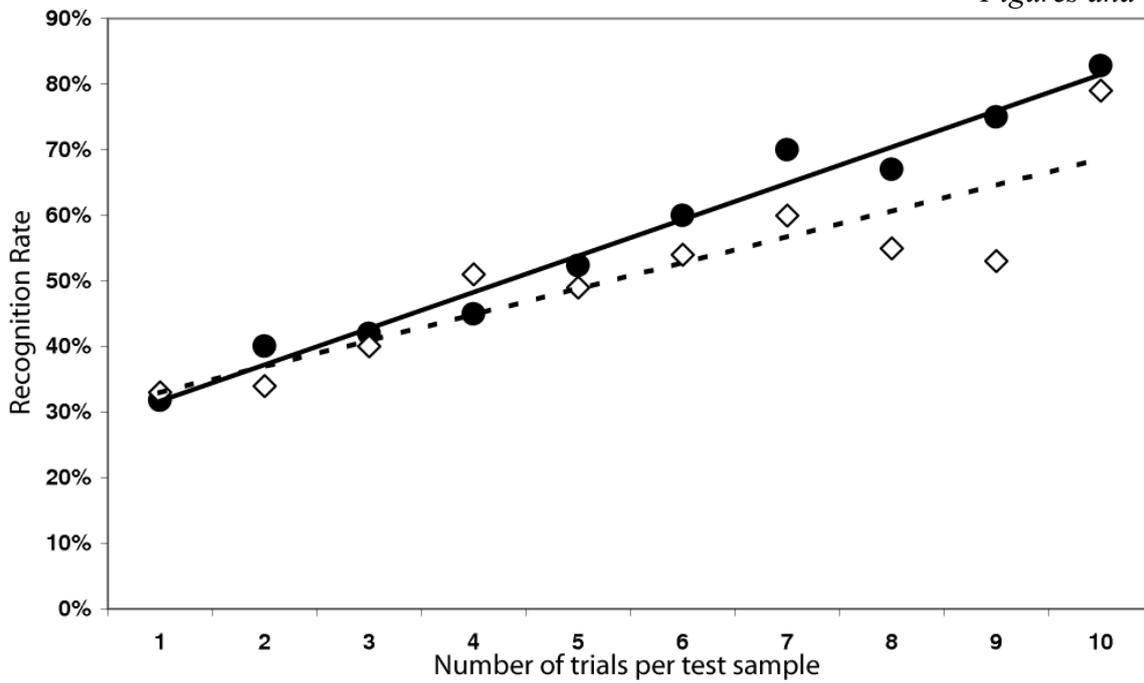

**Figure 3:** Recognition rate for different number of trials per test sample using the best filter for subject S6. The full circle represents recognition rates for the Laplacian, and the black line the best straight line fitting this data. The diamonds are the rates for the potential, with the dashed line representing their best linear fit. Both data are for OE configurations. The Laplacian corresponds to the position for electrode C4A and the potential to electrode C3.

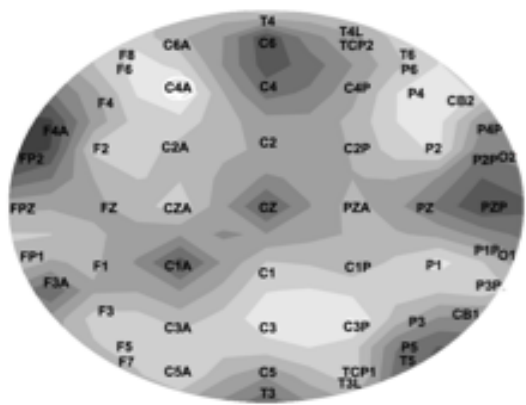
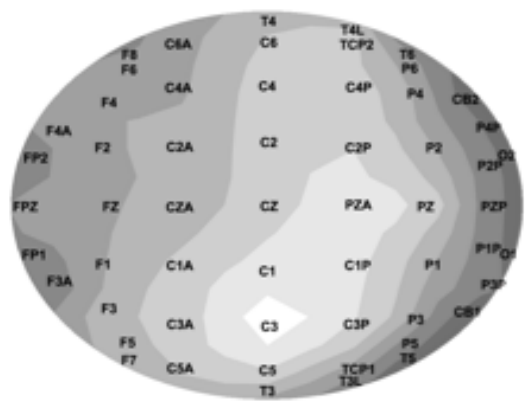

(a) Subject S6, third-order Laplacian interpolation, OE configuration and bandpass [4.5 Hz , 6.5 Hz].

(b) Subject S6, potential analysis, OE configuration, bandpass [2.0 Hz ,12.5 Hz].

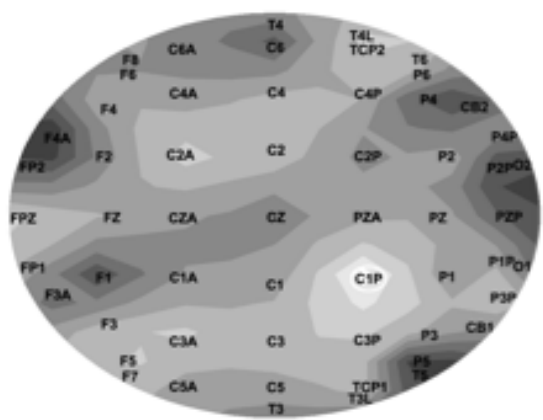
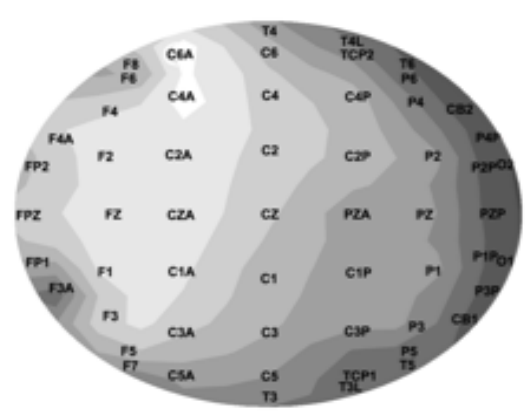

(c) Subject S7, fourth-order Laplacian interpolation, EO configuration and bandpass [1 Hz , 4.5 Hz].

(d) Subject S7, potential analysis, EO configuration and bandpass [2.5 Hz , 10 Hz].

**Figure 4:** Distribution of recognition rates on the scalp for subject S6 and S7, considering the optimal filters showed in Table 1. The lighter areas show better recognition rates.

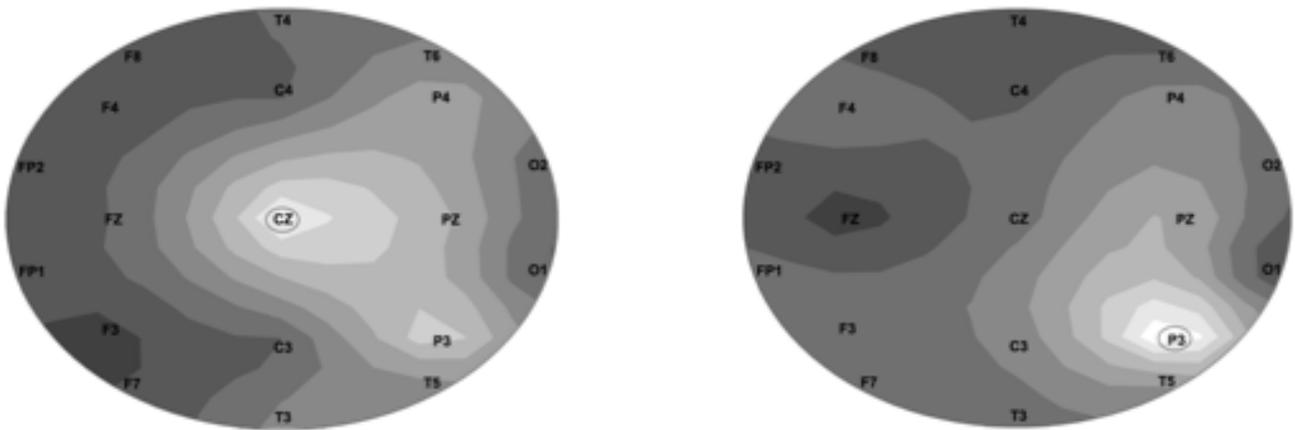

(a) Subject S6, fifth-order Laplacian interpolation, EO configuration and bandpass [4.5 Hz , 22.5 Hz].

(b) Subject S7, fifh-order Laplacian interpolation, EO configuration and bandpass [5 Hz , 9 Hz].

**Figure 5:** Distribution of recognition rates considering optimal filters for 10-20 system.